\documentclass[aps,pra,showpacs,floatfix,twocolumn,amssymb]{revtex4-1}

\usepackage{amsmath}
\usepackage{mathrsfs}
\usepackage{graphicx}
\usepackage{dcolumn}
\usepackage{bm}
\usepackage{euscript}

\def\ket#1{{|#1\rangle}}
\def\bra#1{{\langle#1|}}

\def\cg(#1,#2)(#3,#4)(#5,#6){\bra{#1,#2,#3,#4}#5,#6\rangle}
\def\threej(#1,#2)(#3,#4)(#5,#6){\begin{pmatrix}#1&#3&#5\\#2&#4&#6\end{pmatrix}}
\def\sixj(#1,#2,#3)(#4,#5,#6){\begin{Bmatrix}#1&#2&#3\\#4&#5&#6\end{Bmatrix}}
\def\ninej(#1,#2,#3)(#4,#5,#6)(#7,#8,#9){\begin{Bmatrix}#1&#2&#3\\#4&#5&#6\\#7&#8&#9\end{Bmatrix}}

\begin{document}

\pagenumbering{arabic}

\setcounter{footnote}{0}

\title{Detection of the Meissner Effect with a Diamond Magnetometer}


\author{Louis-S. Bouchard}
 \email{bouchard@chem.ucla.edu}
\affiliation{Department of Chemistry and Biochemistry,  California NanoSystems Institute, Biomedical Engineering IDP, Jonsson Comprehensive Cancer Center, University of California, Los Angeles, CA 90095}

\author{Victor M. Acosta}
 \affiliation{Department of Physics, University of California, Berkeley, California 94720-7300}

\author{Erik Bauch}
 \affiliation{Department of Physics, University of California, Berkeley, California 94720-7300}

\author{Dmitry Budker}
 \affiliation{Department of Physics, University of California, Berkeley, California 94720-7300}

\date{\today}
\begin{abstract}
We examine the possibility of probing superconductivity effects in metal nanoclusters via diamond magnetometry.  Metal nanoclusters have been proposed as constitutive elements of high T$_c$ superconducting nanostructured materials. Magnetometry based on the detection of spin-selective fluorescence of nitrogen-vacancy (NV) centers in diamond  is capable of nanoscale spatial resolution and could be used as a tool for investigating the properties of single or multiple clusters interacting among each other or with a surface.  We have carried out sensitivity estimates and experiments to understand how these magnetometers could be used in such a situation.  We studied the behavior of the sensor as a function of temperature and detected the flux exclusion effect in a superconductor by monitoring the [111]-orientation fine structure spectrum in high-NV density diamond.  Our results show that phase transitions can be ascertained in a bulk superconductor with this technique while the principal zero field splitting parameter ($D$) demonstrates a temperature dependence that may require compensation schemes. 
\end{abstract}



\maketitle


Superconductivity occurs in a wide variety of materials when the temperature drops below a critical temperature T$_c$ and is accompanied by diamagnetism and a drop in resistance to electric current.  The superconducting transitions observed to date are all below 135 K under normal atmospheric pressure and they are mostly associated with bulk systems.  In recent years, the search for room-temperature superconductivity has been approached from several different angles~\cite{bib:rtsc1}, including searching for systems with negative dielectric functions~\cite{bib:rtsc2}, synthesizing crystals with specific phonon spectra~\cite{bib:rtsc3}, creating materials with a pore structure~\cite{bib:rtsc4}, and studying the superconducting state at interfaces in epitaxial heterostructures~\cite{bib:rtsc5}.   

Metal nanoclusters, which can exhibit superconductivity effects~\cite{bib:mcrev1,bib:mcrev2,bib:mcrev3}, also form a novel family and could potentially develop into room-temperature superconductors. Size-selected metal nanoclusters, in the case of certain metals, are known to exhibit electronic shell structure~\cite{bib:mcrev1,bib:mcrev2,bib:mcrev3}.   Kresin and Ovchinnikov recently proposed that a superconducting transition can occur for some metal nanoclusters with 10$^2$-10$^3$ valence electrons~\cite{bib:kresin,bib:KresinReview}. The transitions are expected to occur near electronic shell closings.  Under special circumstances, the shell structure is expected to strengthen pairing correlations, and leads to elevated values for the critical temperature for a transition to the superconducting state. The superconducting state of small nanoclusters is directly related to the phenomenon of pair correlation.  Recently, Cao {\it et al.} have presented experimental measurements of a jump in the heat capacity of Al$_{45}^-$ and Al$_{47}^-$ clusters, which is strong evidence of a superconducting transition in isolated clusters~\cite{bib:caopaper}.  The transition temperature was T$_c \approx$ 200 K.   A number of interesting predictions for other clusters have been made in~\cite{bib:kresin,bib:KresinReview}.

The clusters' exceedingly small sizes (1 nm or less) make them difficult to study in the laboratory, as they are  inaccessible to a large number of common spectroscopies and methods for detecting superconductivity.  It is important to have additional means to probe such clusters at the individual level and in populations, so that cluster-cluster or cluster-surface interactions can be studied.  Recent developments in solid-state magnetometers such as anisotropic magnetoresistive sensors~\cite{bib:amr6,bib:amr5} or those based on diamonds with nitrogen-vacancy (NV) color centers~\cite{bib:gopi1,bib:gopi2,bib:maze} have opened up new possibilities for sensitive magnetic measurements of material properties over a broad range of temperatures with high spatial resolution.  ``NV-diamond" magnetometers have been developed which can perform measurements at the nanoscale, potentially, with single-spin detection~\cite{bib:gopi1,bib:maze}. 

In this article, we look at the feasibility of performing magnetic measurements near the surface of a metal cluster and bulk superconductor to detect the phase transition through the Meissner effect or flux exclusion.  There are three common approaches to finding the critical temperature in superconductors.  These consist of measurements of magnetic susceptibility but also electrical resistivity and specific-heat.  Resistivity measurements at the nanoscale are challenging.  Very often, more than one type of experimental measurement is needed to convincingly prove the existence and probe the nature of the superconducting state.  For heat capacity measurements in individual isolated clusters we note the recent work of Jarrold {\it et al.}~\cite{bib:cao,bib:breaux}.  Susceptibility measurements report changes in the magnetic moment below T$_c$ by way of a susceptometer instrument.  The measurement directly probes the Meissner effect in which a transition to diamagnetism occurs.


Diamagnetism in metal clusters is different from bulk metals.  Clusters containing a magic number of electrons (i.e. filled shell) are diamagnetic regardless of the pairing correlation because of the spherical symmetry of the electronic shell structure~\cite{bib:kresin}.  This is an important deviation from bulk superconductors which indicates there would be no detectable Meissner effect.  For clusters with incomplete shells, however, there is orbital paramagnetism at high temperatures~\cite{bib:kresin} and the transition to the superconducting state at T=T$_c$ should be accompanied by the paramagnetic-diamagnetic transition in the cluster moment.  This suggests there may be a detectable Meissner effect.  On the other hand, the London penetration depth for bulk superconductors (typical values are between 50 nm and 500 nm), is generally much larger than the cluster sizes recently proposed for room-temperature superconductivity~\cite{bib:KresinReview}.  It is therefore unclear if a Meissner effect could be observed via magnetometry experiments in single clusters, or below which cluster size it would become unobservable.  What is clear, however, is that some clusters generally behave differently than their bulk counterpart.

In cluster crystals or suspensions of clusters in an embedding medium, the situation is likely to be different from that of single clusters because of proximity effects.  In Ref.~\cite{bib:heath}, for example, the Meissner effect was observed for Pb nanoparticles embedded in an organic matrix.  Alternative methods of detection which could be implemented by diamond magnetometry include measurements of the Knight shift or spin relaxation times, both of which are known to be indicators of the superconducting phase transition.  For example, T$_1$ exhibits the Slichter-Hebel peak below T$_c$~\cite{bib:hebel}.   T$_1$ could also be used, in favorable cases, to report on the phonon spectrum (see, e.g., Ref.~\cite{Veg2006}) and this would be useful to probe electron-phonon coupling mechanisms.


The development of bulk room-temperature superconductors out of atomic or molecular clusters may require a three-dimensional assembly of clusters into a lattice structure.  Such three-dimensional assembly into cluster crystals has been demonstrated~\cite{bib:gaclustercrystal}.  The study of individual clusters may be best performed by immobilizing individual clusters on a surface.  Recently, ``softlanding" of Ag$_{561}$ clusters on a C$_{60}$ monolayer has been demonstrated~\cite{bib:softlanding}.  Another group has succeeded in immobilizing clusters on a surface~\cite{bib:cluster1} by accurate placement and dimensional control in the assembly of clusters.  This method would enable precise control of the number and position of clusters assembled on a surface, leading to a more systematic way to setup experiments that probe cluster-cluster interactions.  

Magnetic moments of atomic clusters are usually measured in beam experiments using Stern-Gerlach deflection~\cite{bib:yin}.   An interesting variant would use surface magnetic measurements to probe the cluster while it interacts with the surface or in the presence of an embedding material. Probing single clusters rather than populations would enable us to study heterogeneity in the properties among an ensemble of clusters rather than measuring ensemble-averaged properties.  In addition to superconductivity and magnetism, one could also probe insulator-to-metal transitions as a function of cluster size and temperature.  As atoms are added to the cluster, the number of delocalized electrons increases and the cluster's initially paramagnetic state fluctuates and eventually becomes diamagnetic. 


The magnetic properties (magnetization, $M$)  of magnetic materials are often measured with a susceptometer under an applied field ($B$).  The $M-B$ curve is typically probed using a SQUID (superconducting quantum interference device) gradiometer oscillated sinusoidally near the sample and cycling the applied field.  For clusters, a similar kind of experiment may be possible in which a bias field generates a strong enough magnetic polarization $M$ to observe the diamagnetism/paramagnetism of the cluster at the sensor's location.  Let us consider the example of a Na$_{92}$ cluster ($N=92$) placed in a magnetic field, $B=1$~T.  Assuming spherical clusters, the magnetic moment of a diamagnetic cluster is given by the formula~\cite{bib:kittelbook}

\begin{equation}
 \mu_{dia} = - \frac{ e^2 B }{ 6 m } N \langle R^2 \rangle 
\end{equation}

\noindent where $R=a_0 r_s N^{1/3} \approx 9.32 \times 10^{-10}$ m for Na$_{92}$.  $r_s$ is the Wigner-Seitz radius ($r_s=3.9$ for sodium) and $a_0$ is the Bohr radius.  $m$ and $e$ are the mass and charge of the electron, respectively. This leads to $\mu_{dia}=-3.8 \times 10^{-25}$ A$\cdot$m$^2$.  If the sensor is placed at a distance of, for example, 10 nm from the cluster, the induced field to be measured for the diamagnetic state is

\begin{equation}
 B_{dia}  \approx  \frac{ \mu_0}{ 4\pi} \frac{ \mu_{dia} }{ r^3} \approx - 3.8 \times 10^{-8} \mbox{T}. 
\end{equation}

\noindent or $B_{dia} \approx 3.8 \times 10^{-5}$~T at 1~nm distance.  

In the paramagnetic state, we can apply the formula of Pauli paramagnetism for a free electron gas which gives~\cite{bib:kittelbook}

\begin{equation}
 M = \frac{ n \mu_B^2 }{E_F} B,
\end{equation}

\noindent where $n$ in this formula is the number of atoms per unit volume, namely $n=0.97$~mol$/22.989~$cm$^3$ for sodium meta, $\mu_B$ is the Bohr magneton and the Fermi energy for Na is $E_F=k_B T_F = 3.24$~eV.  

This gives a magnetization of $M=\chi_{para} B \approx 0.7 \times 10^{-5}$ T in a 1~T applied field.   The corresponding dipole moment is 

\begin{equation}
\mu_{para}=M \cdot V \approx 1.94 \times 10^{-26}  \mbox{A} \cdot \mbox{m}^2
\end{equation}

\noindent  (radius of Na$_{92}$ cluster is $9 \times 10^{-10}$~m).  The corresponding dipole field is

\begin{equation}
B \approx \frac{ \mu_0 }{ 4\pi} \frac{ \mu }{ r^3 } \approx 2 \times 10^{-9} \mbox{T} 
\end{equation}

\noindent at 10~nm separation distance, or $B\approx 2 \times 10^{-6}$~T at 1~nm distance.   To measure a field of magnitude $10^{-9}$~T with a sensor whose noise floor is 4~nT/$\sqrt{Hz}$ requires about 8 seconds of averaging.  We note that although the paramagnetism is less than the diamagnetism, the phase transition is predicted to be accompanied by a shift from paramagnetic to diamagnetic states upon crossing T$_c$.  This is only true of clusters.  Indeed, diamagnetism in metal clusters is different from bulk metals.  A Meissner effect may be observable in clusters with incomplete shells, due to paramagnetism at high temperatures~\cite{bib:kresin}, and the transition to the superconducting state should be accompanied by the paramagnetic-diamagnetic transition in the cluster moment.  

\begin{figure}[h!]
\includegraphics[width=3.25in]{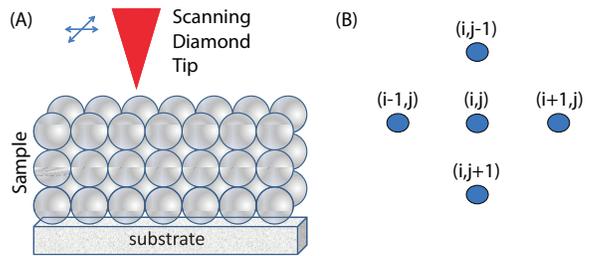}
\caption{\label{fig:proposals} (A)  Experimental setup for surface diamond magnetometry of clusters.  The diamond is scanned across the surface and surface magnetism is measured by subtracting the background signal from neighboring points (B).  (Or equivalently, the sample can be translated rather than the tip.)  }
\end{figure}

A number of practical issues arise for NV-based sensors when a large (1 T) external field is applied.  The three main considerations are: 1) the high microwave frequencies required, 2) the field drift and 3) the background magnetism of the substrate.   The first element has been successfully addressed by Neumann {\it et al.}~\cite{bib:neumann} who have interrogated NV centers in diamond under a bias field of 0.65 T using $\sim$10 GHz microwave fields. The field drift can be mitigated by avoiding fields that are too high and unstable.  The 0.65 T field of~\cite{bib:neumann} is sufficiently stable for optically-detected magnetic resonance (ODMR) measurements. The background magnetism of the substrate can be mitigated by translating the measuring tip near the cluster and performing subtraction of the signal from neighboring points (see Fig.~\ref{fig:proposals}B)~\cite{bib:gopi1}.  This technique has been successfully implemented by Moler's group in a scanning probe micro-SQUID susceptometer~\cite{bib:moler}.  Finally, we note that the energy level structure of the NV center is fundamentally different at high fields ($>$ 0.1 T).  However, the level structure itself should not present a problem for magnetometry.


A possible experimental setup for interrogating atomic clusters and cluster crystals is shown in Fig.~\ref{fig:proposals}A.  A nanoscale diamond tip is positioned near the surface of the cluster crystal or immobilized clusters and the tip is scanned along the surface to provide spatially-resolved information.    The magnetic field gradient can be measured by scanning the tip across the surface and performing background subtractions using neighboring points (Fig.~\ref{fig:proposals}B).  An equivalent scheme is to translate the sample rather than than tip.  Nanodiamonds with a single NV center could be used, but so could nanocrystals containing multiple NV centers.  For detecting superconducting phase transitions, we envisage an experiment in which both the external field and the sample temperature are cycled.

As the cluster size increases the properties approach those of the bulk.  A bulk material consisting of embedded superconducting clusters may exhibit properties that are closer to those of a bulk superconducting material in the sense that diamagnetic effects can be enhanced compared to those of the individual clusters.   Strong fields may not be needed to generate an observable magnetic moment if diamagnetism is created by persistent currents.   In this case, the Meissner effect can be detected using a small external field.  

\begin{figure}
\includegraphics[width=3.25in]{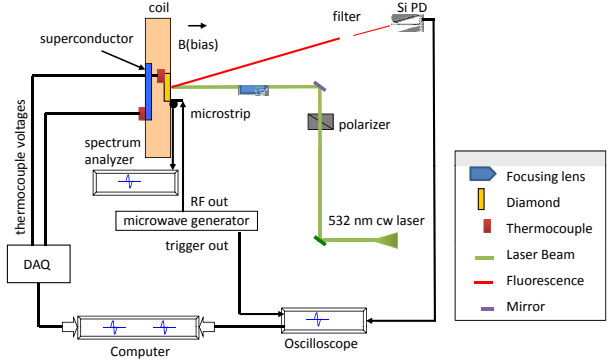}
\caption{\label{fig:setup} 
Schematic diagram of the experimental setup.}
\end{figure}

We now look at ways to measure the Meissner diamagnetism in a weak field using a diamond magnetometer.  The experimental setup is shown in Figure~\ref{fig:setup}. Green light (532 nm) from a 200-mW continuous-wave solid state laser is incident normal to the surface of a [111]-oriented diamond crystal of dimensions 2 mm $\times$ 2 mm $\times$ 1 mm.  The diamond was fabricated commercially by Sumitomo by the high-pressure, high-temperature (HPHT) method with an initial concentration of nitrogen impurities of less than 100 ppm, irradiated with a dose of 10$^{19}$ cm$^{-2}$, 3.0-MeV electrons, then annealed for 2 hours at 700 $^\circ$C~\cite{bib:acosta}.  The fluorescence emitted by the diamond is recorded using a Si photodiode. A 675 $\pm$ 75 nm interference filter is used to attenuate the green light backscattered towards the photodiode.  The BSCCO sample  (BSCCO-2223, Bi$_2$Sr$_2$Ca$_2$Cu$_3$O$_{10}$), a 2.8 cm-diameter, 4 mm thick disc, is placed at a distance of appoximately 5 mm behind the diamond and parallel to the 2 mm $\times$ 2 mm  face of the diamond.  This particular sample-diamond geometric arrangement results in the largest magnetic-field shift due to the Meissner effect.  An ambient bias field was generated by a coil that surrounded the experimental setup and produced $\sim$14 G at the diamond's location.

The diamond was glued to a thin glass slide which was mounted on a printed circuit board with microwave striplines. A ~200-micron diameter copper wire ran in between striplines $\sim$1 mm from the surface of the diamond and delivered cw microwave field ($\sim$13 dBm power, 2.895 to 2.915 GHz frequency-sweep range, 0.20-s sweep repetition time) from a Hewlett Packard 8350B Sweep Oscillator with a HP 83525B RF Plug-In module. The photocurrent from the Si photodiode was recorded using an oscilloscope triggered upon each microwave frequency sweep. A Labview interface was used to download the data from the oscilloscope and record thermocouple voltages at regular intervals.  Two Type-T thermocouples were used to measure temperature.  The first thermocouple was placed within 1 mm of the diamond.  The second was held in close contact with the BSCCO sample, and silicon-based heatsink compound was used to improve thermal contact between the thermocouple and the superconductor.  

\begin{figure}
\includegraphics[width=3.25in]{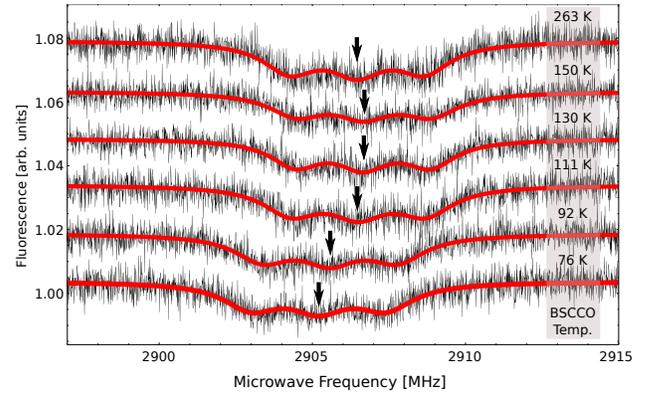}
\caption{\label{fig:111transition} NV-diamond detected microwave spectroscopy of the $\ket{m_S=0} \leftrightarrow \ket{m_S=1}$ transition for different BSCCO temperatures. Three peaks can be resolved which correspond to the hyperfine splitting from the $^{14}$N nucleus.  Depending on the temperature of the BSCCO superconductor, the center of the $\ket{m_S=0} \leftrightarrow \ket{m_S=1}$ transition changes due to two competing effects: the Zeeman shift associated with the Meissner effect and a frequency shift associated with the diamond temperature, which may change as a result of the proximity to the BSCCO. See text for details. }
\end{figure}

Plots of the photocurrent vs. microwave frequency are shown in Fig.~\ref{fig:111transition} for various BSCCO sample temperatures.  For each such graph, the $\ket{m_S=0} \leftrightarrow \ket{m_S=1}$ transition frequency from the [111]-oriented NV centers was extracted by fitting a set of three Lorentzian lines separated by 2.2 MHz corresponding to the hyperfine splitting due to the $^{14}$N nuclei (spin $I=1$), which can be found in three possible orientations $\ket{m_I=0}$, $\ket{m_I=\pm 1}$ with respect to the NV axis~\cite{bib:acosta}.

The $\ket{m_S=0} \leftrightarrow \ket{m_S=1}$ transition center frequency provides a measure of the projection of the local magnetic field $\vec{B}$ along the N-V axis, which causes a frequency shift $g \mu_B \vec{S} \cdot \vec{B}$, where $S=1$ (spin), the magnetogyric ratio $g \mu_B$ is $\approx$2.8 MHz/G for the NV center electronic triplet spin state.  This Zeeman splitting is in addition to the zero-field splitting (ZFS) $H_{cf} = D [S_z^2-(1/3)S(S+1)]$ with $D$ $\approx$ 2.87 GHz along the axis of the NV center (spin $S=1$). For a superconductor below T$_c$ placed in a bias field, we expect the Meissner effect to alter the magnetic field around it, due to the exclusion effect of the superconductor.



\begin{figure}
\includegraphics[width=3.25in]{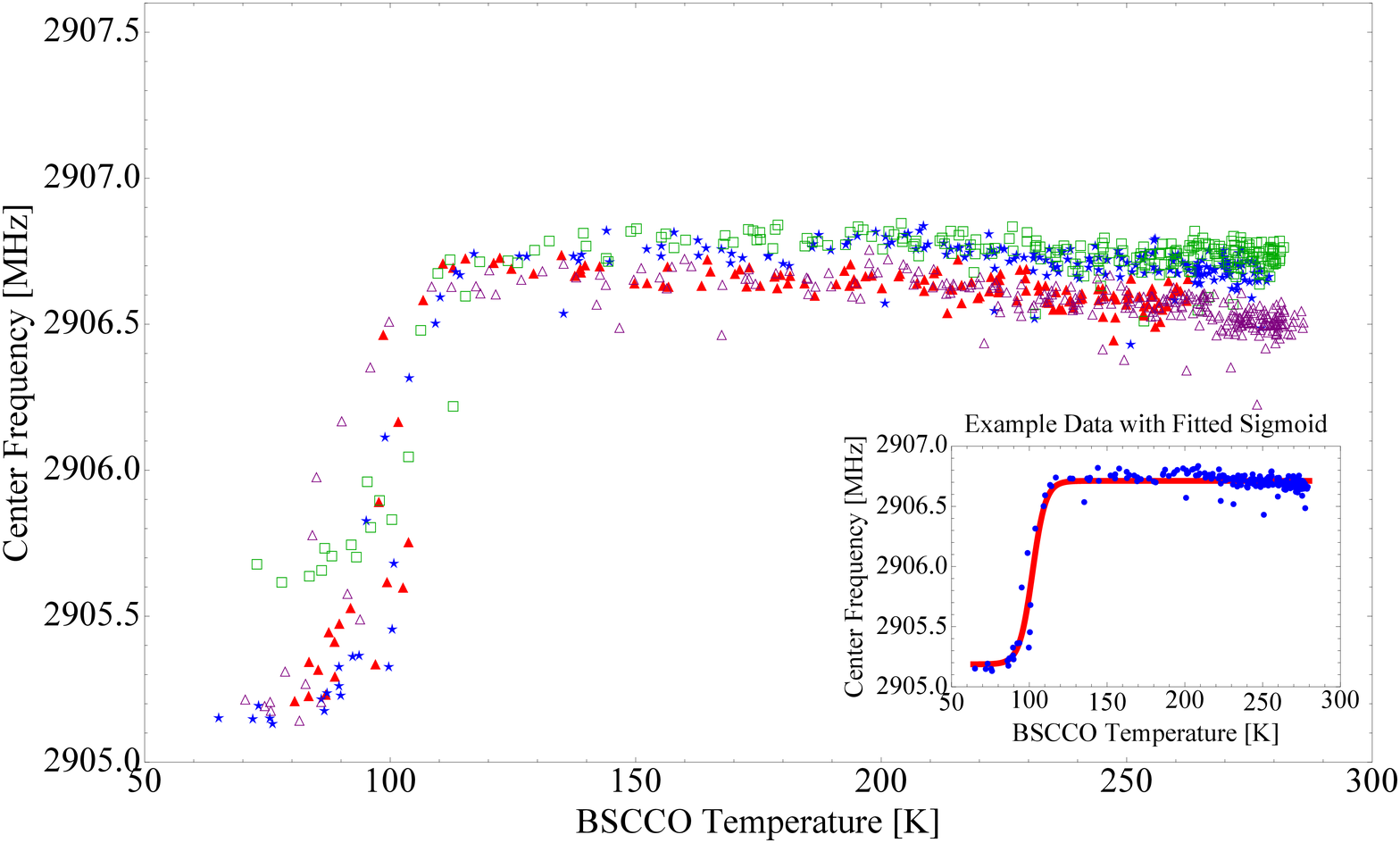}
\caption{\label{fig:bscco} 
BSCCO-2223 shows a phase transition to the superconducting state at T$_c$ of (102 $\pm$ 3) K. The four symbol types represent repeated runs on the same sample perfomed at different times.  }
\end{figure}

Our experiment consists of cooling the BSCCO sample in liquid nitrogen for 30 s, then rapidly positioning it in close proximity (5 mm) to the diamond crystal, with the disc parallel to the crystal plane.  The disc's axis is parallel to $\vec{B}$.  As the BSCCO sample warms up to room-temperature, measurements of the $\ket{m_S=0} \leftrightarrow \ket{m_S=1}$ transition in diamond is measured at regular intervals (every 4 s).  With BSCCO-2223, T$_c$ is 105 K~\cite{bib:bscco}, and this gives us nearly 20 s at ambient temperature to acquire measurements in the presence of the Meissner effect before the phase transition is crossed (T $>$ T$_c$) and the superconductive state is lost.  The results of four different runs performed under similar conditions are shown in Figure~\ref{fig:bscco}. This protocol worked well for BSCCO, but for other materials, a better control of temperature, e.g. using a cryostat, may be required.  

The inset of Figure~\ref{fig:bscco} shows the least squares fit of the data to a sigmoid function

\begin{equation}
f(T) = \frac{f_1}{1+\exp\left[-(T-T_c)/\tau\right]}+f_0
\end{equation}

\noindent for one particular run. The free parameters of the fit are T$_c$, $f_0$ and $f_1$. T$_c$ indicates the phase transition, $f_0$ and $f_1$ are vertical offset and scaling parameters, respectively.  The results for four runs are shown in Table~\ref{tab:1}.  From the fitted value of T$_c$, we determine a standard deviation of 5.7 K.  The average value of T$_c$ is 102 K.

\begin{table}[h]
\begin{tabular}{l*{6}{c}r}
Run        & $f_1$ [MHz] & T$_c$  [K]  & $\tau$ [K] & $f_0$ [MHz]  \\
\hline
1          & 1.3746 & 101.3 & 4.53 & 2905.3  \\
2          & 1.5251 & 102.1 & 4.20 & 2905.2  \\
3          & 1.0883 & 104.3 & 4.71 & 2905.7  \\
4          & 1.4003 & 91.4 & 5.33 & 2905.2  \\
\end{tabular}
\caption{\label{tab:1} Results of the least squares fit of a sigmoid to find the critical temperature. From these data, we find T$_c$ of (102 $\pm$ 3) K. }
\end{table}

Control experiments were performed with a Teflon disc cut to the same dimensions as the BSCCO sample and cooled to liquid nitrogen temperature.  Results (not shown) did not exhibit any signs of a phase transition up to room temperature, as one would expect of a non-superconducting material.  

The data in Figure~\ref{fig:bscco} show a typical spread in the $\ket{m_S=0} \leftrightarrow \ket{m_S=1}$ transition frequency measurements found in our experiments.  This spread is seen to be on the order of ca. 50-100 kHz in the high temperature range.  Part of this drift is caused by drifts of the diamond temperature toward colder temperatures, which we estimate to be at most ca. 10 K, as the cold sample is in close proximity of the diamond. 

While performing the experiments in this study, we have noticed a statistically significant dependence of the ZFS parameter $D$ of the NV center with temperature.  This indicates that extra care should be taken in variable temperature studies, for the effects of $dD/dT$ are otherwise indistinguishable from changes in the external field in the scheme we have used here.  The problem could be mitigated or eliminated either by subtraction of the effects of a temperature drift from independent measurements of temperature, or by using gradiometer schemes.  We have subsequently investigated the temperature dependence in details~\cite{bib:acostaprl} and found a value of $dD/dT=-74.2(7)$ kHz/K in the temperature range 280-330 K.  In particular, we proposed another method to eliminate the effects of temperature drift of the ZFS coefficients based on pulsed ESR (electron spin resonance).


In conclusion, we have demonstrated the use of NV-diamond magnetometry for probing Meissner effects in a bulk superconductor and identified some issues associated with such experiments.  When positioned in close proximity to the surface of a superconductor, changes in the surface field can be detected with temperature.  Care must be taken to ensure that the temperature dependence of the ZFS parameters does not adversely affect the measurement.

We also estimate that it should be possible to probe magnetism in individual metal clusters. Nanoscale diamonds with NV color centers present an opportunity for the study of magnetism and superconductivity in nanomaterials, where sufficient resolution and sensitivity to observe properties of individual clusters could be obtained.  In bulk superconductors, nanoscale NV-diamond magnetometry could be used to image vortices, paramagnetism and nucleation effects and could present advantages over micro-SQUID scanning techniques~\cite{bib:microsquid} by enabling variable temperature and hysteresis studies while maintaining close proximity to the sample throughout the cycle.  Most importantly, high temperatures are inaccessible to SQUIDs without the use of thermal insulation layers that would prevent short-distance operation.  While nucleation effects in the paramagnetic Meissner effect are known to depend on cooling rate~\cite{bib:microsquid} and most likely exhibit hysteresis effects, magnetic imaging studies have not been carried out as a function of temperature across the phase transition.  Such studies could shed light on the origins of the paramagnetism.  Finally, the study of individual metal clusters and their interaction with environment and surfaces could shed light on their roles in chemical catalysis.




This research has been supported by the Director of the Office of Science of the U.S. Department of Energy (through the Nuclear Sciences Divisions of LBNL). This work was supported by a NSF grant (PHY-0855552), an ONR MURI program, and a Dreyfus grant. Useful discussions with V.Z. Kresin and V.V. Kresin are acknowledged.


\bibliography{dbase15}

\begin{thebibliography}{32}%
\makeatletter
\providecommand \@ifxundefined [1]{%
 \@ifx{#1\undefined}
}%
\providecommand \@ifnum [1]{%
 \ifnum #1\expandafter \@firstoftwo
 \else \expandafter \@secondoftwo
 \fi
}%
\providecommand \@ifx [1]{%
 \ifx #1\expandafter \@firstoftwo
 \else \expandafter \@secondoftwo
 \fi
}%
\providecommand \natexlab [1]{#1}%
\providecommand \enquote  [1]{``#1''}%
\providecommand \bibnamefont  [1]{#1}%
\providecommand \bibfnamefont [1]{#1}%
\providecommand \citenamefont [1]{#1}%
\providecommand \href@noop [0]{\@secondoftwo}%
\providecommand \href [0]{\begingroup \@sanitize@url \@href}%
\providecommand \@href[1]{\@@startlink{#1}\@@href}%
\providecommand \@@href[1]{\endgroup#1\@@endlink}%
\providecommand \@sanitize@url [0]{\catcode `\\12\catcode `\$12\catcode
  `\&12\catcode `\#12\catcode `\^12\catcode `\_12\catcode `\%12\relax}%
\providecommand \@@startlink[1]{}%
\providecommand \@@endlink[0]{}%
\providecommand \url  [0]{\begingroup\@sanitize@url \@url }%
\providecommand \@url [1]{\endgroup\@href {#1}{\urlprefix }}%
\providecommand \urlprefix  [0]{URL }%
\providecommand \Eprint [0]{\href }%
\@ifxundefined \urlstyle {%
  \providecommand \doi  [0]{\begingroup \@sanitize@url \@doi}%
  \providecommand \@doi [1]{\endgroup \@@startlink {\doibase
  #1}doi:\discretionary {}{}{}#1\@@endlink }%
}{%
  \providecommand \doi  [0]{doi:\discretionary{}{}{}\begingroup
  \urlstyle{rm}\Url }%
}%
\providecommand \doibase [0]{http://dx.doi.org/}%
\providecommand \Doi [0]{\begingroup \@sanitize@url \@Doi }%
\providecommand \@Doi  [1]{\endgroup\@@startlink{\doibase#1}\@@Doi}%
\providecommand \@@Doi [1]{#1\@@endlink}%
\providecommand \selectlanguage [0]{\@gobble}%
\providecommand \bibinfo  [0]{\@secondoftwo}%
\providecommand \bibfield  [0]{\@secondoftwo}%
\providecommand \translation [1]{[#1]}%
\providecommand \BibitemOpen [0]{}%
\providecommand \bibitemStop [0]{}%
\providecommand \bibitemNoStop [0]{.\EOS\space}%
\providecommand \EOS [0]{\spacefactor3000\relax}%
\providecommand \BibitemShut  [1]{\csname bibitem#1\endcsname}%
\bibitem [{\citenamefont {Ginzburg}(2005)}]{bib:rtsc1}%
  \BibitemOpen
  \bibfield  {author} {\bibinfo {author} {\bibfnamefont {V.}~\bibnamefont
  {Ginzburg}},\ }\href@noop {} {\bibfield  {journal} {\bibinfo  {journal}
  {Phys. Usp.},\ }\textbf {\bibinfo {volume} {48}},\ \bibinfo {pages} {173}
  (\bibinfo {year} {2005})}\BibitemShut {NoStop}%
\bibitem [{\citenamefont {Chu}\ \emph {et~al.}(2005)\citenamefont {Chu},
  \citenamefont {Chen}, \citenamefont {Shulman}, \citenamefont {Tsui},
  \citenamefont {Xue}, \citenamefont {Wen},\ and\ \citenamefont
  {Sheng}}]{bib:rtsc2}%
  \BibitemOpen
  \bibfield  {author} {\bibinfo {author} {\bibfnamefont {C.}~\bibnamefont
  {Chu}}, \bibinfo {author} {\bibfnamefont {F.}~\bibnamefont {Chen}}, \bibinfo
  {author} {\bibfnamefont {J.}~\bibnamefont {Shulman}}, \bibinfo {author}
  {\bibfnamefont {S.}~\bibnamefont {Tsui}}, \bibinfo {author} {\bibfnamefont
  {Y.}~\bibnamefont {Xue}}, \bibinfo {author} {\bibfnamefont {W.}~\bibnamefont
  {Wen}}, \ and\ \bibinfo {author} {\bibfnamefont {P.}~\bibnamefont {Sheng}},\
  }\href@noop {} {\bibfield  {journal} {\bibinfo  {journal}
  {arXiv:cond-mat/0511166v1}} (\bibinfo {year} {2005})}\BibitemShut {NoStop}%
\bibitem [{\citenamefont {Pickett}(2006)}]{bib:rtsc3}%
  \BibitemOpen
  \bibfield  {author} {\bibinfo {author} {\bibfnamefont {W.}~\bibnamefont
  {Pickett}},\ }\href@noop {} {\bibfield  {journal} {\bibinfo  {journal} {J.
  Supercond.},\ }\textbf {\bibinfo {volume} {19}},\ \bibinfo {pages} {291}
  (\bibinfo {year} {2006})}\BibitemShut {NoStop}%
\bibitem [{\citenamefont {Zakhidov}\ and\ \citenamefont
  {et~al.}(1998)}]{bib:rtsc4}%
  \BibitemOpen
  \bibfield  {author} {\bibinfo {author} {\bibfnamefont {A.}~\bibnamefont
  {Zakhidov}}\ and\ \bibinfo {author} {\bibnamefont {et~al.}},\ }\href@noop {}
  {\bibfield  {journal} {\bibinfo  {journal} {Science},\ }\textbf {\bibinfo
  {volume} {282}},\ \bibinfo {pages} {897} (\bibinfo {year}
  {1998})}\BibitemShut {NoStop}%
\bibitem [{\citenamefont {Reyren}\ and\ \citenamefont
  {et~al.}(2007)}]{bib:rtsc5}%
  \BibitemOpen
  \bibfield  {author} {\bibinfo {author} {\bibfnamefont {N.}~\bibnamefont
  {Reyren}}\ and\ \bibinfo {author} {\bibnamefont {et~al.}},\ }\href@noop {}
  {\bibfield  {journal} {\bibinfo  {journal} {Science},\ }\textbf {\bibinfo
  {volume} {317}},\ \bibinfo {pages} {1196} (\bibinfo {year}
  {2007})}\BibitemShut {NoStop}%
\bibitem [{\citenamefont {de~Heer}(1993)}]{bib:mcrev1}%
  \BibitemOpen
  \bibfield  {author} {\bibinfo {author} {\bibfnamefont {W.}~\bibnamefont
  {de~Heer}},\ }\href@noop {} {\bibfield  {journal} {\bibinfo  {journal} {Rev.
  Mod. Phys.},\ }\textbf {\bibinfo {volume} {65}},\ \bibinfo {pages} {611}
  (\bibinfo {year} {1993})}\BibitemShut {NoStop}%
\bibitem [{\citenamefont {Kresin}\ and\ \citenamefont
  {Knight}(1998)}]{bib:mcrev2}%
  \BibitemOpen
  \bibfield  {author} {\bibinfo {author} {\bibfnamefont {V.}~\bibnamefont
  {Kresin}}\ and\ \bibinfo {author} {\bibfnamefont {W.}~\bibnamefont
  {Knight}},\ }\href@noop {} {\bibfield  {journal} {\bibinfo  {journal} {Z.
  Phys. Chem.},\ }\textbf {\bibinfo {volume} {203}},\ \bibinfo {pages} {67}
  (\bibinfo {year} {1998})}\BibitemShut {NoStop}%
\bibitem [{\citenamefont {Brack}(1993)}]{bib:mcrev3}%
  \BibitemOpen
  \bibfield  {author} {\bibinfo {author} {\bibfnamefont {M.}~\bibnamefont
  {Brack}},\ }\href@noop {} {\bibfield  {journal} {\bibinfo  {journal} {Rev.
  Mod. Phys.},\ }\textbf {\bibinfo {volume} {65}},\ \bibinfo {pages} {677}
  (\bibinfo {year} {1993})}\BibitemShut {NoStop}%
\bibitem [{\citenamefont {Ovchinnikov}\ and\ \citenamefont
  {Kresin}(2005)}]{bib:kresin}%
  \BibitemOpen
  \bibfield  {author} {\bibinfo {author} {\bibfnamefont {Y.}~\bibnamefont
  {Ovchinnikov}}\ and\ \bibinfo {author} {\bibfnamefont {V.}~\bibnamefont
  {Kresin}},\ }\href@noop {} {\bibfield  {journal} {\bibinfo  {journal} {Eur.
  Phys. J. B},\ }\textbf {\bibinfo {volume} {45}},\ \bibinfo {pages} {5}
  (\bibinfo {year} {2005})}\BibitemShut {NoStop}%
\bibitem [{\citenamefont {Kresin}\ and\ \citenamefont
  {Ovchinnikov}(2008)}]{bib:KresinReview}%
  \BibitemOpen
  \bibfield  {author} {\bibinfo {author} {\bibfnamefont {V.}~\bibnamefont
  {Kresin}}\ and\ \bibinfo {author} {\bibfnamefont {Y.}~\bibnamefont
  {Ovchinnikov}},\ }\href@noop {} {\bibfield  {journal} {\bibinfo  {journal}
  {Phys. - Uspekhi},\ }\textbf {\bibinfo {volume} {51}},\ \bibinfo {pages}
  {427} (\bibinfo {year} {2008})}\BibitemShut {NoStop}%
\bibitem [{\citenamefont {Cao}\ \emph {et~al.}(2008){\natexlab{a}}\citenamefont
  {Cao}, \citenamefont {Neal}, \citenamefont {Starace}, \citenamefont
  {Ovchinnikov}, \citenamefont {Kresin},\ and\ \citenamefont
  {Jarrold}}]{bib:caopaper}%
  \BibitemOpen
  \bibfield  {author} {\bibinfo {author} {\bibfnamefont {B.}~\bibnamefont
  {Cao}}, \bibinfo {author} {\bibfnamefont {C.}~\bibnamefont {Neal}}, \bibinfo
  {author} {\bibfnamefont {A.}~\bibnamefont {Starace}}, \bibinfo {author}
  {\bibfnamefont {Y.}~\bibnamefont {Ovchinnikov}}, \bibinfo {author}
  {\bibfnamefont {V.}~\bibnamefont {Kresin}}, \ and\ \bibinfo {author}
  {\bibfnamefont {M.}~\bibnamefont {Jarrold}},\ }\href@noop {} {\bibfield
  {journal} {\bibinfo  {journal} {J. Supercond. Nov. Magn.},\ }\textbf
  {\bibinfo {volume} {21}},\ \bibinfo {pages} {163} (\bibinfo {year}
  {2008}{\natexlab{a}})}\BibitemShut {NoStop}%
\bibitem [{\citenamefont {{Sergeeva-Chollet}}\ \emph
  {et~al.}(2010)\citenamefont {{Sergeeva-Chollet}}, \citenamefont {Dyvorne},
  \citenamefont {Polovy}, \citenamefont {{Pannetier-Lecoeur}},\ and\
  \citenamefont {Fermon}}]{bib:amr6}%
  \BibitemOpen
  \bibfield  {author} {\bibinfo {author} {\bibfnamefont {N.}~\bibnamefont
  {{Sergeeva-Chollet}}}, \bibinfo {author} {\bibfnamefont {H.}~\bibnamefont
  {Dyvorne}}, \bibinfo {author} {\bibfnamefont {H.}~\bibnamefont {Polovy}},
  \bibinfo {author} {\bibfnamefont {M.}~\bibnamefont {{Pannetier-Lecoeur}}}, \
  and\ \bibinfo {author} {\bibfnamefont {C.}~\bibnamefont {Fermon}},\ }in\
  \href@noop {} {\emph {\bibinfo {booktitle} {IFMBE Proceedings}}},\ \bibinfo
  {series} {17TH INTERNATIONAL CONFERENCE ON BIOMAGNETISM ADVANCES IN
  BIOMAGNETISM – BIOMAG2010}, Vol.~\bibinfo {volume} {28},\ \bibinfo {editor}
  {edited by\ \bibinfo {editor} {\bibfnamefont {S.}~\bibnamefont {Supek}}\ and\
  \bibinfo {editor} {\bibfnamefont {A.}~\bibnamefont {Susac}}}\ (\bibinfo
  {publisher} {Springer},\ \bibinfo {year} {2010})\ pp.\ \bibinfo {pages}
  {70--73}\BibitemShut {NoStop}%
\bibitem [{\citenamefont {Dyvorne}\ \emph {et~al.}(2009)\citenamefont
  {Dyvorne}, \citenamefont {Fermon}, \citenamefont {{Pannetier-Lecoeur}},
  \citenamefont {Polovy},\ and\ \citenamefont {Walliang}}]{bib:amr5}%
  \BibitemOpen
  \bibfield  {author} {\bibinfo {author} {\bibfnamefont {H.}~\bibnamefont
  {Dyvorne}}, \bibinfo {author} {\bibfnamefont {C.}~\bibnamefont {Fermon}},
  \bibinfo {author} {\bibfnamefont {M.}~\bibnamefont {{Pannetier-Lecoeur}}},
  \bibinfo {author} {\bibfnamefont {H.}~\bibnamefont {Polovy}}, \ and\ \bibinfo
  {author} {\bibfnamefont {A.}~\bibnamefont {Walliang}},\ }\href@noop {}
  {\bibfield  {journal} {\bibinfo  {journal} {IEEE Trans. Appl. Supercon.},\
  }\textbf {\bibinfo {volume} {19}},\ \bibinfo {pages} {819} (\bibinfo {year}
  {2009})}\BibitemShut {NoStop}%
\bibitem [{\citenamefont {Balasubramanian}\ \emph {et~al.}(2008)\citenamefont
  {Balasubramanian}, \citenamefont {Chan}, \citenamefont {Kolesov},
  \citenamefont {Al-Hmoud}, \citenamefont {Tisler}, \citenamefont {Shin},
  \citenamefont {Kim}, \citenamefont {Wojcik}, \citenamefont {Hemmer},
  \citenamefont {Krueger}, \citenamefont {Hanke}, \citenamefont
  {Leitenstorfer}, \citenamefont {Bratschitsch}, \citenamefont {Jelezko},\ and\
  \citenamefont {Wrachtrup}}]{bib:gopi1}%
  \BibitemOpen
  \bibfield  {author} {\bibinfo {author} {\bibfnamefont {G.}~\bibnamefont
  {Balasubramanian}}, \bibinfo {author} {\bibfnamefont {I.}~\bibnamefont
  {Chan}}, \bibinfo {author} {\bibfnamefont {R.}~\bibnamefont {Kolesov}},
  \bibinfo {author} {\bibfnamefont {M.}~\bibnamefont {Al-Hmoud}}, \bibinfo
  {author} {\bibfnamefont {J.}~\bibnamefont {Tisler}}, \bibinfo {author}
  {\bibfnamefont {C.}~\bibnamefont {Shin}}, \bibinfo {author} {\bibfnamefont
  {C.}~\bibnamefont {Kim}}, \bibinfo {author} {\bibfnamefont {A.}~\bibnamefont
  {Wojcik}}, \bibinfo {author} {\bibfnamefont {P.}~\bibnamefont {Hemmer}},
  \bibinfo {author} {\bibfnamefont {A.}~\bibnamefont {Krueger}}, \bibinfo
  {author} {\bibfnamefont {T.}~\bibnamefont {Hanke}}, \bibinfo {author}
  {\bibfnamefont {A.}~\bibnamefont {Leitenstorfer}}, \bibinfo {author}
  {\bibfnamefont {R.}~\bibnamefont {Bratschitsch}}, \bibinfo {author}
  {\bibfnamefont {F.}~\bibnamefont {Jelezko}}, \ and\ \bibinfo {author}
  {\bibfnamefont {J.}~\bibnamefont {Wrachtrup}},\ }\href@noop {} {\bibfield
  {journal} {\bibinfo  {journal} {Nature},\ }\textbf {\bibinfo {volume}
  {455}},\ \bibinfo {pages} {648} (\bibinfo {year} {2008})}\BibitemShut
  {NoStop}%
\bibitem [{\citenamefont {Balasubramanian}\ \emph {et~al.}(2009)\citenamefont
  {Balasubramanian}, \citenamefont {Neumann}, \citenamefont {Twitchen},
  \citenamefont {Markham}, \citenamefont {Kolesov}, \citenamefont {Mizuochi},
  \citenamefont {Isoya}, \citenamefont {Achard}, \citenamefont {Beck},
  \citenamefont {Tissler}, \citenamefont {Jacques}, \citenamefont {Hemmer},
  \citenamefont {Jelezko},\ and\ \citenamefont {Wrachtrup}}]{bib:gopi2}%
  \BibitemOpen
  \bibfield  {author} {\bibinfo {author} {\bibfnamefont {G.}~\bibnamefont
  {Balasubramanian}}, \bibinfo {author} {\bibfnamefont {P.}~\bibnamefont
  {Neumann}}, \bibinfo {author} {\bibfnamefont {D.}~\bibnamefont {Twitchen}},
  \bibinfo {author} {\bibfnamefont {M.}~\bibnamefont {Markham}}, \bibinfo
  {author} {\bibfnamefont {R.}~\bibnamefont {Kolesov}}, \bibinfo {author}
  {\bibfnamefont {N.}~\bibnamefont {Mizuochi}}, \bibinfo {author}
  {\bibfnamefont {J.}~\bibnamefont {Isoya}}, \bibinfo {author} {\bibfnamefont
  {J.}~\bibnamefont {Achard}}, \bibinfo {author} {\bibfnamefont
  {J.}~\bibnamefont {Beck}}, \bibinfo {author} {\bibfnamefont {J.}~\bibnamefont
  {Tissler}}, \bibinfo {author} {\bibfnamefont {V.}~\bibnamefont {Jacques}},
  \bibinfo {author} {\bibfnamefont {P.}~\bibnamefont {Hemmer}}, \bibinfo
  {author} {\bibfnamefont {F.}~\bibnamefont {Jelezko}}, \ and\ \bibinfo
  {author} {\bibfnamefont {J.}~\bibnamefont {Wrachtrup}},\ }\href@noop {}
  {\bibfield  {journal} {\bibinfo  {journal} {Nat. Mat.},\ }\textbf {\bibinfo
  {volume} {8}},\ \bibinfo {pages} {383} (\bibinfo {year} {2009})}\BibitemShut
  {NoStop}%
\bibitem [{\citenamefont {Maze}\ \emph {et~al.}(2008)\citenamefont {Maze},
  \citenamefont {Stanwix}, \citenamefont {Hodges}, \citenamefont {Hong},
  \citenamefont {Taylor}, \citenamefont {Cappellaro}, \citenamefont {Jiang},
  \citenamefont {Gurudev~Dutt}, \citenamefont {Togan}, \citenamefont {Zibrov},
  \citenamefont {Yacoby}, \citenamefont {Walsworth},\ and\ \citenamefont
  {Lukin}}]{bib:maze}%
  \BibitemOpen
  \bibfield  {author} {\bibinfo {author} {\bibfnamefont {J.}~\bibnamefont
  {Maze}}, \bibinfo {author} {\bibfnamefont {P.}~\bibnamefont {Stanwix}},
  \bibinfo {author} {\bibfnamefont {J.}~\bibnamefont {Hodges}}, \bibinfo
  {author} {\bibfnamefont {S.}~\bibnamefont {Hong}}, \bibinfo {author}
  {\bibfnamefont {J.}~\bibnamefont {Taylor}}, \bibinfo {author} {\bibfnamefont
  {P.}~\bibnamefont {Cappellaro}}, \bibinfo {author} {\bibfnamefont
  {L.}~\bibnamefont {Jiang}}, \bibinfo {author} {\bibfnamefont
  {M.}~\bibnamefont {Gurudev~Dutt}}, \bibinfo {author} {\bibfnamefont
  {E.}~\bibnamefont {Togan}}, \bibinfo {author} {\bibfnamefont
  {A.}~\bibnamefont {Zibrov}}, \bibinfo {author} {\bibfnamefont
  {A.}~\bibnamefont {Yacoby}}, \bibinfo {author} {\bibfnamefont
  {R.}~\bibnamefont {Walsworth}}, \ and\ \bibinfo {author} {\bibfnamefont
  {M.}~\bibnamefont {Lukin}},\ }\href@noop {} {\bibfield  {journal} {\bibinfo
  {journal} {Nature},\ }\textbf {\bibinfo {volume} {455}},\ \bibinfo {pages}
  {644} (\bibinfo {year} {2008})}\BibitemShut {NoStop}%
\bibitem [{\citenamefont {Cao}\ \emph {et~al.}(2008){\natexlab{b}}\citenamefont
  {Cao}, \citenamefont {Neal}, \citenamefont {Starace}, \citenamefont
  {Ovchinnikov}, \citenamefont {Kresin},\ and\ \citenamefont
  {Jarrold}}]{bib:cao}%
  \BibitemOpen
  \bibfield  {author} {\bibinfo {author} {\bibfnamefont {B.}~\bibnamefont
  {Cao}}, \bibinfo {author} {\bibfnamefont {C.~M.}\ \bibnamefont {Neal}},
  \bibinfo {author} {\bibfnamefont {A.~K.}\ \bibnamefont {Starace}}, \bibinfo
  {author} {\bibfnamefont {Y.~N.}\ \bibnamefont {Ovchinnikov}}, \bibinfo
  {author} {\bibfnamefont {V.~Z.}\ \bibnamefont {Kresin}}, \ and\ \bibinfo
  {author} {\bibfnamefont {M.~F.}\ \bibnamefont {Jarrold}},\ }\href@noop {}
  {\bibfield  {journal} {\bibinfo  {journal} {arXiv:0804.0824}} (\bibinfo
  {year} {2008}{\natexlab{b}})}\BibitemShut {NoStop}%
\bibitem [{\citenamefont {Breaux}\ and\ \citenamefont
  {et~al.}(2005)}]{bib:breaux}%
  \BibitemOpen
  \bibfield  {author} {\bibinfo {author} {\bibfnamefont {G.}~\bibnamefont
  {Breaux}}\ and\ \bibinfo {author} {\bibnamefont {et~al.}},\ }\href@noop {}
  {\bibfield  {journal} {\bibinfo  {journal} {Phys. Rev. Lett.},\ }\textbf
  {\bibinfo {volume} {94}},\ \bibinfo {pages} {173401} (\bibinfo {year}
  {2005})}\BibitemShut {NoStop}%
\bibitem [{\citenamefont {Weitz}\ \emph {et~al.}(2000)\citenamefont {Weitz},
  \citenamefont {Sample}, \citenamefont {Ries}, \citenamefont {Spain},\ and\
  \citenamefont {Heath}}]{bib:heath}%
  \BibitemOpen
  \bibfield  {author} {\bibinfo {author} {\bibfnamefont {I.}~\bibnamefont
  {Weitz}}, \bibinfo {author} {\bibfnamefont {J.}~\bibnamefont {Sample}},
  \bibinfo {author} {\bibfnamefont {R.}~\bibnamefont {Ries}}, \bibinfo {author}
  {\bibfnamefont {E.}~\bibnamefont {Spain}}, \ and\ \bibinfo {author}
  {\bibfnamefont {J.}~\bibnamefont {Heath}},\ }\href@noop {} {\bibfield
  {journal} {\bibinfo  {journal} {J. Phys. Chem. B},\ }\textbf {\bibinfo
  {volume} {104}},\ \bibinfo {pages} {4288} (\bibinfo {year}
  {2000})}\BibitemShut {NoStop}%
\bibitem [{\citenamefont {Hebel}\ and\ \citenamefont
  {Slichter}(1959)}]{bib:hebel}%
  \BibitemOpen
  \bibfield  {author} {\bibinfo {author} {\bibfnamefont {L.}~\bibnamefont
  {Hebel}}\ and\ \bibinfo {author} {\bibfnamefont {C.}~\bibnamefont
  {Slichter}},\ }\href@noop {} {\bibfield  {journal} {\bibinfo  {journal}
  {Phys. Rev.},\ }\textbf {\bibinfo {volume} {113}},\ \bibinfo {pages} {1504}
  (\bibinfo {year} {1959})}\BibitemShut {NoStop}%
\bibitem [{\citenamefont {Vega}\ \emph {et~al.}(2006)\citenamefont {Vega},
  \citenamefont {Beckmann}, \citenamefont {Bai},\ and\ \citenamefont
  {Dybowski}}]{Veg2006}%
  \BibitemOpen
  \bibfield  {author} {\bibinfo {author} {\bibfnamefont {A.~J.}\ \bibnamefont
  {Vega}}, \bibinfo {author} {\bibfnamefont {P.~A.}\ \bibnamefont {Beckmann}},
  \bibinfo {author} {\bibfnamefont {S.}~\bibnamefont {Bai}}, \ and\ \bibinfo
  {author} {\bibfnamefont {C.}~\bibnamefont {Dybowski}},\ }\href@noop {}
  {\bibfield  {journal} {\bibinfo  {journal} {Phys. Rev. B},\ }\textbf
  {\bibinfo {volume} {74}},\ \bibinfo {pages} {214420} (\bibinfo {year}
  {2006})}\BibitemShut {NoStop}%
\bibitem [{\citenamefont {Hagel}\ \emph {et~al.}(2002)\citenamefont {Hagel},
  \citenamefont {Kelemen}, \citenamefont {Fischer}, \citenamefont {Pilawa},
  \citenamefont {Wisnitza}, \citenamefont {Dormann}, \citenamefont
  {L{\"{o}}hneysen}, \citenamefont {Schnepf}, \citenamefont {Schn{\"{o}}ckel},
  \citenamefont {Neisel},\ and\ \citenamefont {Beck}}]{bib:gaclustercrystal}%
  \BibitemOpen
  \bibfield  {author} {\bibinfo {author} {\bibfnamefont {J.}~\bibnamefont
  {Hagel}}, \bibinfo {author} {\bibfnamefont {M.}~\bibnamefont {Kelemen}},
  \bibinfo {author} {\bibfnamefont {G.}~\bibnamefont {Fischer}}, \bibinfo
  {author} {\bibfnamefont {B.}~\bibnamefont {Pilawa}}, \bibinfo {author}
  {\bibfnamefont {J.}~\bibnamefont {Wisnitza}}, \bibinfo {author}
  {\bibfnamefont {E.}~\bibnamefont {Dormann}}, \bibinfo {author} {\bibfnamefont
  {H.}~\bibnamefont {L{\"{o}}hneysen}}, \bibinfo {author} {\bibfnamefont
  {A.}~\bibnamefont {Schnepf}}, \bibinfo {author} {\bibfnamefont
  {H.}~\bibnamefont {Schn{\"{o}}ckel}}, \bibinfo {author} {\bibfnamefont
  {U.}~\bibnamefont {Neisel}}, \ and\ \bibinfo {author} {\bibfnamefont
  {J.}~\bibnamefont {Beck}},\ }\href@noop {} {\bibfield  {journal} {\bibinfo
  {journal} {J. Low Temp. Phys.},\ }\textbf {\bibinfo {volume} {129}},\
  \bibinfo {pages} {133} (\bibinfo {year} {2002})}\BibitemShut {NoStop}%
\bibitem [{\citenamefont {Duffe}\ \emph {et~al.}(2007)\citenamefont {Duffe},
  \citenamefont {Irawan}, \citenamefont {Bieletzki}, \citenamefont {Richter},
  \citenamefont {Sieben}, \citenamefont {Yin}, \citenamefont {von Issendorff},
  \citenamefont {Moseler},\ and\ \citenamefont {H{\"o}vel}}]{bib:softlanding}%
  \BibitemOpen
  \bibfield  {author} {\bibinfo {author} {\bibfnamefont {S.}~\bibnamefont
  {Duffe}}, \bibinfo {author} {\bibfnamefont {T.}~\bibnamefont {Irawan}},
  \bibinfo {author} {\bibfnamefont {M.}~\bibnamefont {Bieletzki}}, \bibinfo
  {author} {\bibfnamefont {T.}~\bibnamefont {Richter}}, \bibinfo {author}
  {\bibfnamefont {B.}~\bibnamefont {Sieben}}, \bibinfo {author} {\bibfnamefont
  {C.}~\bibnamefont {Yin}}, \bibinfo {author} {\bibfnamefont {B.}~\bibnamefont
  {von Issendorff}}, \bibinfo {author} {\bibfnamefont {M.}~\bibnamefont
  {Moseler}}, \ and\ \bibinfo {author} {\bibfnamefont {H.}~\bibnamefont
  {H{\"o}vel}},\ }\href@noop {} {\bibfield  {journal} {\bibinfo  {journal}
  {Eur. Phys. J. D},\ }\textbf {\bibinfo {volume} {45}},\ \bibinfo {pages}
  {401} (\bibinfo {year} {2007})}\BibitemShut {NoStop}%
\bibitem [{\citenamefont {Partridge}\ \emph {et~al.}(2006)\citenamefont
  {Partridge}, \citenamefont {Reichel}, \citenamefont {Ayesh}, \citenamefont
  {Mackenzie},\ and\ \citenamefont {Brown}}]{bib:cluster1}%
  \BibitemOpen
  \bibfield  {author} {\bibinfo {author} {\bibfnamefont {J.}~\bibnamefont
  {Partridge}}, \bibinfo {author} {\bibfnamefont {R.}~\bibnamefont {Reichel}},
  \bibinfo {author} {\bibfnamefont {A.}~\bibnamefont {Ayesh}}, \bibinfo
  {author} {\bibfnamefont {D.}~\bibnamefont {Mackenzie}}, \ and\ \bibinfo
  {author} {\bibfnamefont {S.}~\bibnamefont {Brown}},\ }\href@noop {}
  {\bibfield  {journal} {\bibinfo  {journal} {Phys. Stat. Sol. (a)},\ }\textbf
  {\bibinfo {volume} {203}},\ \bibinfo {pages} {1217} (\bibinfo {year}
  {2006})}\BibitemShut {NoStop}%
\bibitem [{\citenamefont {Yin}\ \emph {et~al.}(2007)\citenamefont {Yin},
  \citenamefont {Moro}, \citenamefont {Xu},\ and\ \citenamefont {{de
  Heer}}}]{bib:yin}%
  \BibitemOpen
  \bibfield  {author} {\bibinfo {author} {\bibfnamefont {S.}~\bibnamefont
  {Yin}}, \bibinfo {author} {\bibfnamefont {R.}~\bibnamefont {Moro}}, \bibinfo
  {author} {\bibfnamefont {X.}~\bibnamefont {Xu}}, \ and\ \bibinfo {author}
  {\bibfnamefont {W.}~\bibnamefont {{de Heer}}},\ }\href@noop {} {\bibfield
  {journal} {\bibinfo  {journal} {Phys. Rev. Lett.},\ }\textbf {\bibinfo
  {volume} {98}},\ \bibinfo {pages} {113401} (\bibinfo {year}
  {2007})}\BibitemShut {NoStop}%
\bibitem [{\citenamefont {Kittel}(2004)}]{bib:kittelbook}%
  \BibitemOpen
  \bibfield  {author} {\bibinfo {author} {\bibfnamefont {C.}~\bibnamefont
  {Kittel}},\ }\href@noop {} {\emph {\bibinfo {title} {Introduction to solid
  state physics}}},\ \bibinfo {edition} {8th}\ ed.\ (\bibinfo  {publisher}
  {Wiley},\ \bibinfo {address} {New York},\ \bibinfo {year} {2004})\BibitemShut
  {NoStop}%
\bibitem [{\citenamefont {Neumann}\ \emph {et~al.}(2010)\citenamefont
  {Neumann}, \citenamefont {Beck}, \citenamefont {Steiner}, \citenamefont
  {Rempp}, \citenamefont {Fedder}, \citenamefont {Hemmer}, \citenamefont
  {Wrachtrup},\ and\ \citenamefont {Jelezko}}]{bib:neumann}%
  \BibitemOpen
  \bibfield  {author} {\bibinfo {author} {\bibfnamefont {P.}~\bibnamefont
  {Neumann}}, \bibinfo {author} {\bibfnamefont {J.}~\bibnamefont {Beck}},
  \bibinfo {author} {\bibfnamefont {M.}~\bibnamefont {Steiner}}, \bibinfo
  {author} {\bibfnamefont {F.}~\bibnamefont {Rempp}}, \bibinfo {author}
  {\bibfnamefont {H.}~\bibnamefont {Fedder}}, \bibinfo {author} {\bibfnamefont
  {P.}~\bibnamefont {Hemmer}}, \bibinfo {author} {\bibfnamefont
  {J.}~\bibnamefont {Wrachtrup}}, \ and\ \bibinfo {author} {\bibfnamefont
  {F.}~\bibnamefont {Jelezko}},\ }\href@noop {} {\bibfield  {journal} {\bibinfo
   {journal} {Science},\ }\textbf {\bibinfo {volume} {329}},\ \bibinfo {pages}
  {542} (\bibinfo {year} {2010})}\BibitemShut {NoStop}%
\bibitem [{\citenamefont {Huber}\ \emph {et~al.}(2008)\citenamefont {Huber},
  \citenamefont {Koshnick}, \citenamefont {Bluhm}, \citenamefont {Archuleta},
  \citenamefont {Azua}, \citenamefont {Bjornsson}, \citenamefont {Gardner},
  \citenamefont {Halloran}, \citenamefont {Lucero},\ and\ \citenamefont
  {Moler}}]{bib:moler}%
  \BibitemOpen
  \bibfield  {author} {\bibinfo {author} {\bibfnamefont {M.}~\bibnamefont
  {Huber}}, \bibinfo {author} {\bibfnamefont {N.}~\bibnamefont {Koshnick}},
  \bibinfo {author} {\bibfnamefont {H.}~\bibnamefont {Bluhm}}, \bibinfo
  {author} {\bibfnamefont {L.}~\bibnamefont {Archuleta}}, \bibinfo {author}
  {\bibfnamefont {T.}~\bibnamefont {Azua}}, \bibinfo {author} {\bibfnamefont
  {P.}~\bibnamefont {Bjornsson}}, \bibinfo {author} {\bibfnamefont
  {B.}~\bibnamefont {Gardner}}, \bibinfo {author} {\bibfnamefont
  {S.}~\bibnamefont {Halloran}}, \bibinfo {author} {\bibfnamefont
  {E.}~\bibnamefont {Lucero}}, \ and\ \bibinfo {author} {\bibfnamefont
  {K.}~\bibnamefont {Moler}},\ }\href@noop {} {\bibfield  {journal} {\bibinfo
  {journal} {Rev. Sci. Instrum.},\ }\textbf {\bibinfo {volume} {79}},\ \bibinfo
  {pages} {053704} (\bibinfo {year} {2008})}\BibitemShut {NoStop}%
\bibitem [{\citenamefont {Acosta}\ \emph {et~al.}(2009)\citenamefont {Acosta},
  \citenamefont {Bauch}, \citenamefont {Ledbetter}, \citenamefont {Santori},
  \citenamefont {Fu}, \citenamefont {Barclay}, \citenamefont {Beausoleil},
  \citenamefont {Linget}, \citenamefont {Roch}, \citenamefont {Treussart},
  \citenamefont {Chemerisov}, \citenamefont {Gawlik},\ and\ \citenamefont
  {Budker}}]{bib:acosta}%
  \BibitemOpen
  \bibfield  {author} {\bibinfo {author} {\bibfnamefont {V.}~\bibnamefont
  {Acosta}}, \bibinfo {author} {\bibfnamefont {E.}~\bibnamefont {Bauch}},
  \bibinfo {author} {\bibfnamefont {M.}~\bibnamefont {Ledbetter}}, \bibinfo
  {author} {\bibfnamefont {C.}~\bibnamefont {Santori}}, \bibinfo {author}
  {\bibfnamefont {K.}~\bibnamefont {Fu}}, \bibinfo {author} {\bibfnamefont
  {P.}~\bibnamefont {Barclay}}, \bibinfo {author} {\bibfnamefont
  {R.}~\bibnamefont {Beausoleil}}, \bibinfo {author} {\bibfnamefont
  {H.}~\bibnamefont {Linget}}, \bibinfo {author} {\bibfnamefont
  {J.}~\bibnamefont {Roch}}, \bibinfo {author} {\bibfnamefont {F.}~\bibnamefont
  {Treussart}}, \bibinfo {author} {\bibfnamefont {S.}~\bibnamefont
  {Chemerisov}}, \bibinfo {author} {\bibfnamefont {W.}~\bibnamefont {Gawlik}},
  \ and\ \bibinfo {author} {\bibfnamefont {D.}~\bibnamefont {Budker}},\
  }\href@noop {} {\bibfield  {journal} {\bibinfo  {journal} {Phys. Rev. B},\
  }\textbf {\bibinfo {volume} {80}},\ \bibinfo {pages} {115202} (\bibinfo
  {year} {2009})}\BibitemShut {NoStop}%
\bibitem [{\citenamefont {Lim}\ and\ \citenamefont {Byrne}(1997)}]{bib:bscco}%
  \BibitemOpen
  \bibfield  {author} {\bibinfo {author} {\bibfnamefont {H.}~\bibnamefont
  {Lim}}\ and\ \bibinfo {author} {\bibfnamefont {J.}~\bibnamefont {Byrne}},\
  }\href@noop {} {\bibfield  {journal} {\bibinfo  {journal} {Metallurgical Mat.
  Trans. B},\ }\textbf {\bibinfo {volume} {28}},\ \bibinfo {pages} {425}
  (\bibinfo {year} {1997})}\BibitemShut {NoStop}%
\bibitem [{\citenamefont {Acosta}\ \emph {et~al.}(2010)\citenamefont {Acosta},
  \citenamefont {Bauch}, \citenamefont {Ledbetter}, \citenamefont {Waxman},
  \citenamefont {Bouchard},\ and\ \citenamefont {Budker}}]{bib:acostaprl}%
  \BibitemOpen
  \bibfield  {author} {\bibinfo {author} {\bibfnamefont {V.}~\bibnamefont
  {Acosta}}, \bibinfo {author} {\bibfnamefont {E.}~\bibnamefont {Bauch}},
  \bibinfo {author} {\bibfnamefont {M.}~\bibnamefont {Ledbetter}}, \bibinfo
  {author} {\bibfnamefont {A.}~\bibnamefont {Waxman}}, \bibinfo {author}
  {\bibfnamefont {L.}~\bibnamefont {Bouchard}}, \ and\ \bibinfo {author}
  {\bibfnamefont {D.}~\bibnamefont {Budker}},\ }\href@noop {} {\bibfield
  {journal} {\bibinfo  {journal} {Phys. Rev. Lett.},\ }\textbf {\bibinfo
  {volume} {104}},\ \bibinfo {pages} {070801} (\bibinfo {year}
  {2010})}\BibitemShut {NoStop}%
\bibitem [{\citenamefont {Kirtley}\ \emph {et~al.}(1998)\citenamefont
  {Kirtley}, \citenamefont {Mota}, \citenamefont {Sigrist},\ and\ \citenamefont
  {Rice}}]{bib:microsquid}%
  \BibitemOpen
  \bibfield  {author} {\bibinfo {author} {\bibfnamefont {J.}~\bibnamefont
  {Kirtley}}, \bibinfo {author} {\bibfnamefont {A.}~\bibnamefont {Mota}},
  \bibinfo {author} {\bibfnamefont {M.}~\bibnamefont {Sigrist}}, \ and\
  \bibinfo {author} {\bibfnamefont {T.}~\bibnamefont {Rice}},\ }\href@noop {}
  {\bibfield  {journal} {\bibinfo  {journal} {J. Phys.: Condens. Matter},\
  }\textbf {\bibinfo {volume} {10}},\ \bibinfo {pages} {L97} (\bibinfo {year}
  {1998})}\BibitemShut {NoStop}%
\end{thebibliography}%

\end{document}